\documentclass[11pt]{article}

\pdfoutput=1 

\usepackage{jheppub} 
\usepackage[T1]{fontenc} 
\usepackage{subfig,amsmath,amsfonts,amssymb,yhmath}
\usepackage{bbm}
\usepackage{todonotes}
\usepackage{lmodern}
\usepackage{tikz}
\usetikzlibrary{decorations.pathmorphing}
\usepackage{caption}
\usepackage{subfig}
\usepackage{graphicx}
\usepackage{verbatim}
\usepackage{cleveref}
\usepackage{braket}
\usepackage{array}
\usepackage{makecell}
\tikzset{snake it/.style={decorate, decoration=snake}}
\tikzset{coil it/.style={decorate, decoration={coil,aspect=0.6,segment length=2mm,amplitude=1mm}}}
\def\ie{\begin{equation}\begin{aligned}}
\def\fe{\end{aligned}\end{equation}}

\title{Irregular Fibonacci Conformal Blocks}
\author[\ast]{Xia Gu,}
\author[\ast,\dag]{Babak Haghighat}
\author[\ast]{and Kevin Loo}
\affiliation[\ast]{Yau Mathematical Sciences Center, Tsinghua University, Beijing, 100084, China}
\affiliation[\dag]{Yanqi Lake Beijing Institute of Mathematical Sciences and Applications (BIMSA), Huairou District, Beijing 101408, P. R. China}
\emailAdd{gux19@mails.tsinghua.edu.cn}
\emailAdd{babakhaghighat@tsinghua.edu.cn}
\emailAdd{lu-jy20@mails.tsinghua.edu.cn}
\date{}
\abstract{This work studies Liouville conformal blocks of irregular type with the insertion of at least one level-$3$ degenerate field admitting a Fibonacci fusion rule. We algebraically derive the corresponding third-order BPZ equations for regular blocks and their modifications when a rank one irregular operator is inserted. Employing Lefschetz thimbles as integration cycles, we then successively proceed to construct integral representations and prove that they satisfy the corresponding BPZ equations. Finally, we show that taking a semiclassical limit, these integral representations can be expressed in terms of Heun functions and have correct leading behaviors consistent with conformal weights and fusion rules.}
\begin{document}
\maketitle
\section{Introduction}
Two-dimensional conformal field theories (2D CFT) has been an intensive field of study since the seminal paper \cite{Belavin:1984vu}. The correlation functions of primary fields in 2D CFT are subject to strong symmetry restrictions which is often cast in the form of differential equations. Among all the symmetry-related equations, there are global ones generated by lower modes of the Virasoro algebra $L_0$ and $L_{-1}$ which hold for any primaries. And there are also local constraints available only for the degenerate primaries, meaning those containing null descendants in their conformal family. These degenerate primaries are labeled by a pair of positive integers $(r,s)$ introduced by Kac \cite{Kac:1978ge}. The product of Kac labels, $rs$, then gives the order of the null differential equations satisfied by the corresponding conformal blocks. The simplest case of degenerate fields is the $(1,2)$-field constrained by second-order equations, for which the $\sigma$ Ising spin field is an example.

One systematic way to construct a solution to these differential equations is the \emph{Coulomb gas formalism}. Introduced by Dotsenko and Fateev \cite{Dotsenko:1984nm}, the Coulomb gas formalism constructs solutions of null equations in arbitrary  CFTs as an integration of correlators of free boson vertex operators. The vertex operators carry $U(1)$ charges $\alpha$, which must be neutral in non-vanishing correlators. This puts a selection rule on possible Kac labels appearing in a non-vanishing integral. In a minimal model, of which the number of primary fields are finite, the Kac labels $(r,s)$ existing in the theory are restricted to small $r$ and $s$. On the other hand, the \emph{Liouville theory} is a CFT with infinite number of primaries and continuous spectrum of the conformal dimension, which is also referred to as non-compact CFT. The $(r,s)$-degenerate fields, with arbitrary $r$ and $s$, coexist in a single Liouville theory, which makes the $U(1)$ charge selection rule easier to satisfy.

The continuous spectrum of Liouville theory makes another interesting construction possible, namely the \emph{irregular states}. Irregular states have a defining property \cite{Gaiotto:2012sf} of being the eigenvector of some positive Virasoro modes $L_k$, $k>0$. Acting with negative modes on these irregular states generates a new type of Virasoro module which is beyond the highest-weight module characterization. The irregular states and related constructions are first considered by Gaiotto \cite{Gaiotto:2009ma} in extending the AGT correspondence \cite{Alday:2009aq} to 4d asymptotically free gauge theories of Argyres-Douglas type with higher order (irregular) singularities. On the 2D CFT side, these higher order singularities just appear in the OPE of the energy-momentum tensor $T$ and the irregular fields $I$, and transform into higher order singularities in the symmetry differential equation. This means the conformal block containing irregular fields exhibits novel features related to irregular singularities, such as the Stokes phenomena \cite{Gu_2022} and the appearance of new types of conformal block bundles \cite{haghighat2023flat}. Further revealing the structure of irregular fields and their relation to other fields is an intriguing subject of study.

In this work, we would like to investigate the conformal blocks of the $(1,3)$-field with and without irregular singularities by differential equations and Coulomb gas formalism. It is the second simplest case next to the $(1,2)$-case which is already discussed in \cite{haghighat2023flat,Bonelli_2022}. Our motivation emerges from the following two observations. As we already mentioned, the order of the differential equation concerned is equal to $rs=3$, which is more involved as compared to the $rs=2$ case and provides a closer look at the interaction of irregular states with general $(r,s)$ degenerate fields. 
From another perspective, in the tricritical Ising model $\mathcal{M}(4,5)$, the $(1,3)$ field is the thermal operator $\epsilon'$ which admits a Fibonacci-anyon-like fusion rule. The braiding of Fibonacci anyons was shown \cite{Freedman:modular} to provide a universal set of quantum computing gates. The natural question to ask is whether and how inserting irregular fields in the conformal block would affect this property.

The organization of the paper is as follows. In section 2, we provide the preliminaries about the Liouville theory and the Coulomb gas formalism. In section 3 we derive in detail the differential equation satisfied by the $
(1,3)$-field and give a Coulomb gas integral solution. In this regular case, we also obtain the monodromy behavior using the standard approach \cite{DiFrancesco:639405}. In section 4 we replace one of the regular operators with a rank-1 irregular operator in the conformal block and prove that the Coulomb gas representation and the null vector differential equation can be adapted to this case consistently. We also consider the semiclassical limit of irregular conformal blocks, which is useful in studying the monodromy behavior. Finally in section 5, we present our conclusions and some future directions.

\section{Liouville field theory and degenerate fields}
We start by considering a conformal field theory associated with a finite-dimensional semi-simple complex Lie algebra $\mathfrak{g}$, described by the action functional \cite{Fateev:2007ab,Fateev:2008bm}
\begin{equation}
    \mathcal{S}(\phi,g)= \frac{1}{8\pi}\int_\Sigma d\tau d\sigma \sqrt{g}\left[g^{\mu\nu}\left\langle\partial_\mu\phi,\partial_\nu\phi\right\rangle+8\pi\mu \sum_{j=1}^{r} e^{b \left\langle e_j,\phi\right\rangle}
    +2R\left\langle Q\rho,\phi\right\rangle\right].
\end{equation}
This is known as the \emph{Toda field theory}, defined on a Riemann surface $\Sigma$ with the metric $g_{\mu\nu}\,(\mu,\nu=\sigma\, \text{or}\, \tau)$ and Ricci scalar $R$.
Throughout this work, we consider the background to be the genus-0 Riemann surface, i.e. the unit 2-sphere $\mathbf{S}^2$. Equivalently, this is the extended complex plane $\mathbb{C}\cup \{\infty\}$ equipped with the metric
\begin{equation}
    g=\frac{4}{(1+|z|^2)^2}|dz|^2
\end{equation}
defined via the stereographic projection of a two-sphere $\mathbf{S}^2$. The parameter $b$ serves as a coupling constant and $\mu$ stands for the cosmological constant.

We define the dimension of Cartan subalgebra $\mathfrak{h}$ of $\mathfrak{g}$ to be the rank $r$ of $\mathfrak{g}$.  The set $\{e_j\}_{j=1,\cdots,r}$ constitutes the simple roots of the Lie algebra $\mathfrak{g}$, which spans the dual vector space $\mathfrak{h}^*$. $\rho$ is the Weyl vector of $\mathfrak{h}^*$, which is half the sum of all positive roots.
An alternative basis of $\mathfrak{h}^*$ is the fundamental weight $\omega_i, 1\leq i\leq r$ defined by 
\begin{equation}
    \omega_i:=\sum_{j=1}^r (A^{-1})_{ij} e_j,
\end{equation}
where $A$ is the Cartan matrix of $\mathfrak{g}$ with entries
\begin{equation}
    A_{ij}:=2\frac{\langle e_i,e_j\rangle}{\langle e_i,e_i\rangle}, \quad \forall\, 1\leq i,j\leq r.
\end{equation}
This basis has the following property,
\begin{equation}
    \langle \omega_i,\omega_j\rangle= \sum_{k,k'=1}^r (A^{-1})_{ik} (A)_{kk'}(A^{-1})_{k'j}=(A^{-1})_{ij}.
\end{equation}
In these definitions, $\left\langle\cdot,\cdot\right\rangle$ is the Killing form (canonical pairing) in $\mathfrak{g}$. 
$\phi$ is the scalar field valued in the Cartan subalgebra $\mathfrak{h}$:
\begin{equation}
    \phi=\sum_{i=1}^r \phi_i \omega_i.
\end{equation}
The central charge of the theory can be parameterized as
\begin{equation}
    c=r\left[1+h^\vee(h^\vee+1)Q^2\right],
\end{equation}
with $h^\vee$ the dual Coxeter number of $\mathfrak{g}$, following the Freudenthal-de Vries strange formula (see for instance, \cite{Bernard:1989iy}). The background charge $Q$ is related to $b$ by 
\begin{equation}
    Q=b+b^{-1}.
\end{equation} 
The theory is unitary only for specific values of $Q$ which fit $c$ into the Kac table \cite{DiFrancesco:639405}. This implies that one can choose $b=i\sqrt{\frac{p}{q}}$ where $p$ and $q$ are coprime, i.e. $\operatorname{gcd}(p,q)=1$.

For $\mathfrak{g}=\mathfrak{a}_1=\mathfrak{sl}_2$ we specify the following data:
\begin{equation}
    r=1,\quad h^\vee=2,\quad e=2, \quad c=1+6Q^2.
\end{equation}
These data in turn defines the \emph{Liouville field theory} to be the simplest case of Toda theories with the action functional
\begin{equation}
    S(\phi,g)= \frac{1}{16\pi}\int_\Sigma d\tau d\sigma  \sqrt{g}\left[g^{\mu\nu} \partial_\mu\phi \partial_\nu\phi
    +16\pi\mu e^{2b\phi}
    +2RQ \phi\right].
\end{equation}

The operator algebra of this theory is the tensor product of the infinite-dimensional holomorphic and anti-holomorphic Virasoro algebras $Vir\otimes \overline{Vir}$. We will consider only the holomorphic part for the rest of our discussions.
The Virasoro algebra is an associative algebra generated by the raising and lowering operators $L_n$. These operators are modes in the Laurent expansions of the spin-2 stress tensor $T(z)$:
\begin{equation}
    T(z)=\sum_{n=-\infty}^\infty\frac{L_n}{z^{n+2}}.
\end{equation}
The highest weight representation of the Verma module -- the $Vir$ representation space, identified as the primary field $\Psi_\Delta$, is labelled by the eigenvalue (also known as the conformal weight) $\Delta$ of the zero mode $L_0$:
\begin{equation}
        L_0\Psi =\Delta\Psi, \quad \Delta\in\mathbb{R}_+.
\end{equation}
Other generators of the algebra
$L_{n}$ with $n<0$ create new fields called the descendant fields, while all positive modes $L_n, \, n>0$ annihilate the primary field.
All these modes are the Virasoro generators and satisfy the Virasoro commutation relations
\begin{equation}
    [L_m,L_n] =(m-n)L_{m+n}+\frac{c}{12}m(m^2-1)\delta_{m+n,0}~.
\end{equation}

\subsection{Vertex operators}
Viewing the Liouville theory as an $\mathfrak{sl}_2$ Toda field theory, a (chiral) primary field\footnote{In fact, the primary field is represented by the vertex operator $\Psi_\alpha(z)=N_\alpha^{-1}V_\alpha(z)$, up to a normalization constant $N_\alpha^{-1}$ \cite{Fateev:2001mj}.} of conformal weight $\Delta=\alpha(Q-\alpha)$ can be represented by a normal-ordered free bosonic exponential, known as the (chiral) vertex operator
\begin{equation}
    V_\alpha(z)= \; :\exp{(\langle\alpha,\phi(z)\rangle)}:\;
    = \; :\exp{(2\alpha\phi(z))}:,
\end{equation}
where $\phi$ is the scalar field defined above and the parameter $\alpha$ is referred to as the \emph{momentum}.

Among all the primaries, there is a special set called degenerate primaries, the momenta of which are of the form
\begin{equation} 
    \begin{aligned}
    \alpha &=\frac{(1-r)b}{2}+\frac{1-s}{2b}, \quad r,s\in \mathbb{Z}_{\geq 0}~.
    \end{aligned}
\end{equation}
We denote such a vertex operator corresponding to these degenerate fields as $V(r,s)$. 
For general values of $c$, there are infinite number of degenerate fields. However, it was shown \cite{Fateev:1987vh} that taking $b=i\sqrt{\frac{p+1}{p}}$, the space of degenerate operators constitutes a closed operator algebra\footnote{The choice of $b$ is in accordance with the coprime constraint on $b$, as mentioned in the previous section.}. This is equivalent to restricting the values to
\begin{equation}     
    \begin{aligned}
    1\leq r\leq p-1,\quad 1\leq s\leq p,\quad \text{for}\; p\geq3,
    \end{aligned}
\end{equation}
and we obtain a finite number of local fields which close under fusion. This goes under the name of \emph{minimal model} $\mathcal{M}(p,p+1)$.

Note that the conformal weight $\Delta$ of the degenerate vertex operator can now be written as follows:
\begin{equation} 
    \Delta(r,s) =\frac{[(p+1)r-ps]^2-1}{4(p+1)p}.
\end{equation}
From the above formula, it is obvious that $\Delta$ possesses a symmetry (which corresponds to one of the $\mathfrak{sl}_2$ Weyl reflection symmetries):
\begin{equation}
    \Delta(r,s)=\Delta(p-1-r,p-s).
\end{equation}
The complete actions of the Weyl group on the vertex operator $V(r,s)$ with momentum $(r,s)$ are given by\footnote{Strictly speaking, these vertex operators are proportional to each other, with a proportionality
constant called the reflection coefficient.}
\begin{equation}\label{eq:weylpotts}
V(r,s)=V(-r,-s)=V(p-1-r,p-s)=V(r-p+1,s-p). 
\end{equation}

Since the vertex operators represent the primary fields, when acting with the Virasoro generators on these operators will generate the Verma module consisting of the primary and all its descendants.

\subsection{Coulomb gas formalism}
Recall that the vertex operator takes the form $V_\alpha(z)= \; :\exp{(2\alpha\phi(z))}:$.
The $n$-point correlation function of vertex operators is given by
\begin{equation}
    \left\langle\prod_{i=1}^n V_{\alpha_i}(z_i)\right\rangle=
    \int[\mathcal{D}\phi]e^{-S(\phi,g)} \prod_{i=1}^n V_{\alpha_i}(z_i).
\end{equation}
In the presence of background charge $Q\rho$ ($\rho=1$ for Liouville theory) at infinity, the correlation function of vertex operators should vanish unless the background charge is screened by momenta (or charges) of the vertex operators:
\begin{equation}
     \left\langle\prod_{i=1}^n V_{\alpha_i}(z_i)\right\rangle=
     \begin{cases}
        \prod_{i<j}(z_i-z_j)^{-2\alpha_i\cdot \alpha_j} & \text{if } \sum_{i=1}^n \alpha_i=Q,\\ 
        0 & \text{otherwise. }
    \end{cases}
\end{equation}
Here we have made use of the fact that the correlation function of the free boson $\phi$ is normalized as
\begin{equation}
    \langle \phi(z)\phi(z')\rangle=-\frac12\ln(z-z').
\end{equation}

In general, the screening condition is not naturally satisfied. In order to remedy this situation, we will associate the vertex operator at infinity with momentum $Q-\alpha$ and introduce screening operators of vanishing conformal weights but carrying nonzero momenta. Such operators do not exist in local form, but can be obtained by integrating along curves in $\mathbb{C}$ the exponential fields of conformal weight 1.  
These exponential fields are the screening current $V_{\alpha_+}(w)$ and its dual $V_{\alpha_-}(w)$, with $\alpha_+=b$ and $\alpha_-=b^{-1}$. Hence the screening charges corresponding to these currents are 
\begin{equation}
\begin{aligned}
    \mathcal{Q}_+ &=\int_\gamma dw\, V_{\alpha_+}(w)=\int_\gamma dw\, e^{\alpha_+\phi(w)}=\int_\gamma dw\, e^{b\phi(w)}, \\
    \mathcal{Q}_- &=\int_{\gamma'} dw\, V_{\alpha_-}(w)
    \int_\gamma dw\, e^{\alpha_-\phi(w)}=\int_{\gamma'} dw\, e^{\phi(w)/b},
\end{aligned}
\end{equation}
where $\gamma$ and $\gamma'$ are integration contours, which can either be compact or noncompact. We will use thimbles $\mathcal{J}$, to be defined later, as our integration contours in the following. Taking into account these screening charges, the correlator of $n$ primaries on a complex plane takes the form
\begin{equation}
\label{eq:freerep}
    \left\langle\prod_{i=1}^n \mathcal{O}_{\alpha_i}(z_i)\right\rangle\propto 
    \left\langle \mathcal{Q}_+^p\mathcal{Q}_-^q \prod_{i=1}^n V_{\alpha_i}(z_i)\right\rangle
\end{equation}
for some insertion numbers $p,q\in \mathbb{Z}_{+}$.
The condition for a non-vanishing correlator is now modified to be
\begin{equation}
    \sum_{i=1}^{n}\alpha_i=(1-p)b+\frac{1-q}{b}.
\end{equation}
This restriction is called the charge neutrality condition or the conservation of free field momentum\footnote{A different approach to obtain the charge neutrality condition from the correlator is the so-called zero-mode integration method \cite{Goulian:1990qr}.}.
We note that, acting with elements of the Weyl group on any of the $V_{\alpha_i}$'s may change the total number of screening charges. 
The construction we have so far described is the free field realization of conformal blocks corresponding to the correlation function that is considered. This procedure is known as the \emph{Coulomb gas formalism} in the literature.

\subsection{Lefschetz thimbles}
Motivated by the conformal block at hand (which we will discuss in detail in Section \ref{sec:regcb}), we consider a block with $q$ screening charges of the same momentum $\alpha_-$. Each screening charge contributes to a contour integral over the coordinate $w_i$. 

One can define a logarithmic function such that the conformal block is expressed as an exponential function of $w_i$ and $z_a$,
\begin{equation}\label{eq:yyfun}
   \mathcal{B}(z)
   =  \int_{\mathcal{C}_i}  \exp{\left(\frac{1}{b^2}\mathcal{W}(z_a,w_i)\right)}\prod_{i}dw_i.
\end{equation}
Such a function is known as the Yang-Yang function\footnote{The Yang-Yang function was named after the two authors in the work \cite{Nekrasov:2011bc} relating this function in Hitchin's quantum integrable systems to four-dimensional supersymmetric gauge theory.}  $\mathcal{W}(z_a,w_i)$, which is a function of the holomorphic coordinates $w_i$ and $z_a$ for various $i$'s and $a$'s. In the following, assume our $b$ is a pure imaginary number. Define the Morse function $h$ to be
\begin{equation}
h(z_a,w_i)=\mathfrak{Re}\left[-\mathcal{W}(z_a,w_i)\right],
\end{equation}
To ensure the convergence of the integral, $h$ has to approach $-\infty$ at the noncompact ends of the thimbles, which can either be the points $w_i\to z_a$ or infinity. We can see that $h$ is maximum at the critical points\footnote{The critical points are also known, in literature, as saddle points or stationary points. They are the fixed points of the gradient flow.} $w_*$ of $\mathcal{W}$. 

Critical points are the points at which all the first derivatives of $\mathcal{W}$ vanish, $\partial_{w_i}\mathcal{W}=0$. Setting $h$ as the Morse function means that the critical points are non-degenerate in which the matrix of second derivatives  $\partial^2_{w_i}\mathcal{W}$ is invertible at $w_*$'s.

Instead of assuming the integration cycles $\mathcal{C}_1$ and $\mathcal{C}_2$ to be closed contours -- which are compact cycles used in the usual description of conformal blocks \cite{DiFrancesco:639405} -- we consider noncompact cycles known as the (Lefschetz) thimbles. 

Suppose we have a $\mathcal{W}$ which is a function of a single variable $w$, a thimble is the union of points on the steepest descent paths flowing out from the critical points $w_*$. More precisely, introducing a fictitious ``time'' parameter $t\in (-\infty,0]$, the thimble consists of all the curves $w(t)$ that solve the gradient flow equations
\begin{equation} \label{eq:flow}
\begin{gathered} 
    \frac{d\overline{w}(t)}{dt}=-\frac{\partial \mathcal{W}}{\partial w(t)},\quad
    \frac{d w(t)}{dt}=-\frac{\partial\overline{\mathcal{W}}}{\partial\overline{w}(t)},
\end{gathered}
\end{equation}
with the initial condition $w(-\infty)=w_*$.
Alternatively, separating the real and complex parts of the variable $w$ as $w=\mathfrak{Re}w+i \mathfrak{Im}w$ gives a computable form of flow equations \eqref{eq:flow}:
\begin{equation}
	\label{downward flow}
	\frac{d\,\mathfrak{Re}w(t)}{dt}=-\mathfrak{Re}\left(\frac{\partial\mathcal{W}}{\partial w(t)}\right),\quad \frac{d\,\mathfrak{Im}w(t)}{dt}=\mathfrak{Im}\left(\frac{\partial\mathcal{W}}{\partial w(t)}\right).
\end{equation}
We will use $\mathcal{J}$ to denote the thimble\footnote{Noncompact integration cycles such as thimbles are necessary specifically in the presence of irregular singularity at infinity, since some integration cycles have ends at $w=\infty$. Integrals involving irregular singularity will be discussed further in Section \ref{sec:irreg}.} hereafter.
Note that, the position of the critical points $w_*$ and the thimble $\mathcal{J}$ depend on the value of $z$. Another property is that the integrand goes to $0$ as $w$ approaches asymptotic regions of the thimble. This fact will be important in section 4 when verifying the solution of differential equations. 

We can naturally apply the same construction of thimbles to the Yang-Yang function with multiple variables $w_i, i=1,...,n$. The critical points can be determined, upon taking $\partial_{w_i} \mathcal{W}=0$, by the so-called Bethe equations
\begin{equation}
    \sum_a \frac{1}{w_i-z_a}=\sum_{j\neq i} \frac{1}{w_i-w_j},\quad i=1,...,n.
\end{equation}
 The resulting thimble is a subspace of real dimension $n$ in the $2n$-dimensional domain.

\section{Regular conformal blocks}\label{sec:regcb}
In this section we derive the BPZ-type equation and conformal blocks for higher degenerate fields. In particular we consider the vertex operator $V(1,3)$, which coincides with the thermal operator $\epsilon'$ in the tricritical Ising model. 

\subsection{Level three  null vectors}
A degenerate vertex operator of Kac label $(r,s)=(1,3)$ or $(3,1)$ exhibits a null vector -- which is a descendant operator with zero norm -- at level $rs=3$. One can derive such a null vector by the following ansatz:
\begin{equation}
    \mathcal{N}_\alpha=\left(\lambda_1  L_{-1}^3
    +\lambda_{12}  L_{-1}L_{-2}+\lambda_{21}  L_{-2}L_{-1} +\lambda_{111} L_{-3}\right)V_{\alpha},
\end{equation} 
Since we have $L_3=[L_2,L_1]$, we can simplify $\mathcal{N}_\alpha$ into
\begin{equation}
    \mathcal{N}_\alpha=\left(\lambda_1  L_{-1}^3
    +\lambda_{2}  L_{-1}L_{-2} +\lambda_3 L_{-3}\right)V_{\alpha},
\end{equation}
where $\lambda$'s are constants which will be determined by the constraints $L_n \mathcal{N}_\alpha=0$ for $n\geq 1$.
By utilizing the Virasoro commutation relations, we have 
\begin{equation}\label{eq:nvf}
\begin{aligned}
    0&=L_1 \mathcal{N}_\alpha=\left[ (2\Delta+4)\lambda_2+4\lambda_3\right]L_{-2}V_{\alpha}+\left[ (6\Delta+6)\lambda_1+3\lambda_2\right]L_{-1}^2V_{\alpha},\\
    0&=L_2 \mathcal{N}_\alpha=\left[(18\Delta+6)\lambda_1+\left(4\Delta+9+\frac{c}{2}\right)\lambda_2+5\lambda_3\right]L_{-1}V_{\alpha}.
\end{aligned}
\end{equation}
Here only two relations are needed to determine the constants since $L_3 \mathcal{N}_\alpha$ is a redundant condition in which it can be obtained via the commutator $L_3=[L_2,L_1]$.

Solving \eqref{eq:nvf} gives the level 3 null state via the operator-state correspondence:
\begin{equation}
\label{eq:null1,3}
    \ket{\chi}=  \left((\Delta+2) L_{-3}-2L_{-1}L_{-2}+\frac{1}{\Delta+1}L_{-1}^3\right)\ket{\Delta}=0,
\end{equation}
where the equality to zero means the decoupling of the null state from the Verma module.
Inserting this state in a correlation function with three other arbitrary primaries leads to a differential equation for the correlation function:
\begin{equation}\label{eq:l3null}
\begin{aligned}
   0 &= \braket{\chi(z_1)\prod_{i=2}^4 \mathcal{O}_i(z_i)}\\
   &= \left((\Delta+2)\mathcal{L}_{-3}-2\mathcal{L}_{-1}\mathcal{L}_{-2}+\frac{1}{\Delta+1}\mathcal{L}_{-1}^3\right)  \braket{\Delta(z_1)\prod_{i=2}^4 \mathcal{O}_i(z_i)},
\end{aligned}
\end{equation}
with $\mathcal{L}_{-k}$ being the differential operators
\begin{equation}
\begin{aligned}
    \mathcal{L}_{-k}=\begin{cases}
        \partial_{z_1}, &k=1,\\
        \sum_{i=2}^4\left[\frac{(k-1) \Delta_i}{(z_i-z_1)^k}-\frac{1}{(z_i-z_1)^{k-1}}\partial_{z_i}\right], & k>1.
    \end{cases}
\end{aligned}
\end{equation}
Explicitly, this is a third-order partial differential equation ($\Delta_1$ denotes the conformal weight of $\ket{\Delta}$):
\begin{equation}\label{eq:4pde}
    \begin{aligned}
        &\left[-(\Delta_1+2)\sum_{i=2}^4\left(\frac{2 \Delta_i}{(z_1-z_i)^3}+\frac{1}{(z_1-z_i)^{2}}\partial_{z_i}\right)
        -2 \partial_{z_1} \sum_{i=2}^4\left(\frac{\Delta_i}{(z_1-z_i)^2}+\frac{1}{z_1-z_i}\partial_{z_i}\right)
        +\frac{1}{\Delta_1+1}\partial_{z_1}^3\right]\\  
&\quad \times \braket{\Delta(z_1)\prod_{i=2}^4 \mathcal{O}_i(z_i)}=0
    \end{aligned}
\end{equation}
and using the shorthand notation $z_{ij}\equiv z_i-z_j$, it is given by
\begin{equation}
    \begin{aligned}
        &\left[\frac{1}{\Delta_1+1}\partial_{z_1}^3
        +4\left(\frac{\Delta_2}{(z_{12})^3}+\frac{\Delta_3}{(z_{13})^3}+\frac{\Delta_4}{(z_{14})^3}\right)
        -2 \left(\frac{\Delta_2}{(z_{12})^2}+\frac{\Delta_3}{(z_{13})^2}+\frac{\Delta_4}{(z_{14})^2}\right)\partial_{z_1}\right.\\
        &\quad 
        +2 \left(\frac{1}{z_{12}^2}\partial_{z_2}+\frac{1}{z_{13}^2}\partial_{z_3}+\frac{1}{z_{14}^2}\partial_{z_4}\right)
        -2 \left(\frac{1}{z_{12}}\partial_{z_1}\partial_{z_2}+\frac{1}{z_{13}}\partial_{z_1}\partial_{z_3}+\frac{1}{z_{14}}\partial_{z_1}\partial_{z_4}\right)\\
        &\quad \left.
        -(\Delta_1+2)\left(\frac{2\Delta_2}{(z_{12})^3}
        +\frac{2\Delta_3}{(z_{13})^3}+\frac{2\Delta_4}{(z_{14})^3}
        +\frac{1}{(z_{12})^2}\partial_{z_2}
        +\frac{1}{(z_{13})^2}\partial_{z_3}
        +\frac{1}{(z_{14})^2}\partial_{z_4}\right)
        \right]\\
        &\quad\times \braket{\Delta(z_1)\prod_{i=2}^4 \mathcal{O}_i(z_i)}=0.
    \end{aligned}
\end{equation}

This equation is greatly simplified by employing the global conformal invariance of the correlator, which is generated by the modes $L_{-1}, L_0, L_1$.
This is encoded in the following global conformal Ward identities:
\begin{equation}\label{eq:wardid}
    \begin{aligned}
        L_{-1} \text{ invariance}: &\quad \sum_{j=1}^{4}\partial_{z_j}\braket{\Delta(z_1)\prod_{i=2}^4 \mathcal{O}_i(z_i)}=0,\\
        L_{0} \text{ invariance}: &\quad
        \sum_{j=1}^{4} \left(z_j\partial_{z_j}+\Delta_j\right)\braket{\Delta(z_1)\prod_{i=2}^4 \mathcal{O}_i(z_i)}=0,\\
        L_{1} \text{ invariance}: &\quad 
        \sum_{j=1}^{4} \left(z_j^2\partial_{z_j}+2z_j\Delta_j\right)\braket{\Delta(z_1)\prod_{i=2}^4 \mathcal{O}_i(z_i)}=0.
    \end{aligned}
\end{equation}
These three identities fix the correlator in the form
\begin{equation}
    \braket{\Delta(z_1)\mathcal{O}_1(z_2)\mathcal{O}_2(z_3)\mathcal{O}_3(z_4)}=\mathfrak{L}(\{z_{ij}\})G(z),
\end{equation}
where $G(z)$ is a yet-to-be-determined function of cross-ratio
\begin{equation}
    z=\frac{z_{12}z_{34}}{z_{14}z_{32}},
\end{equation}
and $\mathfrak{L}(\{z_{ij}\})$ is the leg factor containing only the difference in coordinates $z_{ij}=z_i-z_j$:
\begin{equation}
        \mathfrak{L}(\{z_{ij}\})=\prod_{1\leq i<j\leq 4} z_{ij}^{\mu_{ij}},\quad \mu_{ij}=\frac{1}{3}\left(\sum_{k=1}^4 \Delta_k\right)-\Delta_i-\Delta_j,
\end{equation}
provided $\mu_{ij}$ is symmetric in its indices ($\mu_{ij}=\mu_{ji}$) and $\mu_{ij}$ is a solution of
\begin{equation}
    \sum_{i\neq j} \mu_{ij}=2\Delta_i.
\end{equation}
 
With the invariance of the correlator under global conformal transformations \eqref{eq:wardid}, one can consider the limits $z_1\to z, z_2\to 0, z_3\to 1$, and $z_4\to\infty$. The actions of the derivatives appearing in equation \eqref{eq:4pde} are given by
\begin{equation}\begin{aligned}
    \partial_{z_1} (\mathfrak{L} G) &= \mathfrak{L}\left[\frac{\mu_{12}}{z}
    +\frac{\mu_{13}}{z-1}+\partial_z\right] G \\
    \partial_{z_2} (\mathfrak{L} G) &= \mathfrak{L}\left[-\frac{\mu_{12}}{z}-\mu_{23}+(z-1)\partial_z\right] G \\
    \partial_{z_3} (\mathfrak{L} G) &= \mathfrak{L}\left[-\frac{\mu_{13}}{z-1}+\mu_{23}-z\partial_z\right] G \\
    \partial_{z_4} (\mathfrak{L} G) &= 0\\
    \partial_{z_1}^2 (\mathfrak{L} G) 
    &= \mathfrak{L}\left[\frac{\mu_{12}(\mu_{12}-1)}{z^2}
    +\frac{\mu_{13}(\mu_{13}-1)}{(z-1)^2}
    +2\frac{\mu_{12}\mu_{13}}{z(z-1)}
    +2\left( \frac{\mu_{12}}{z}+\frac{\mu_{13}}{z-1}\right)\partial_z+\partial_z^2\right] G \\
    \partial_{z_1}^3 (\mathfrak{L} G) 
    &= \mathfrak{L}\left[\frac{\mu_{12}(\mu_{12}-1)(\mu_{12}-2)}{z^3}
    +\frac{\mu_{13}(\mu_{13}-1)(\mu_{13}-2)}{(z-1)^3}
    \right. \\
    &\quad +\frac{3\mu_{12}(\mu_{12}-1)\mu_{13}}{z^2(z-1)}
    +\frac{3\mu_{12}\mu_{13}(\mu_{13}-1)}{z(z-1)^2}\\
    &\left.\quad +3\left(\frac{\mu_{12}(\mu_{12}-1)}{z^2}
    +\frac{\mu_{13}(\mu_{13}-1)}{(z-1)^2}+\frac{2\mu_{12}\mu_{13}}{z(z-1)}\right)\partial_z
    +3\left(\frac{\mu_{12}}{z}+\frac{\mu_{13}}{z-1}\right)\partial_z^2
    + \partial_z^3\right] G.\\
\end{aligned}\end{equation}
Plugging in these operators leads to a differential equation for $G(z)$. One can obtain the differential equation for the correlator by inserting back the leg factor as follows:
\begin{equation}
    H(z)\equiv\braket{\Delta(z)\mathcal{O}_1(0)\mathcal{O}_2(1)\mathcal{O}_3(\infty)}=z^{\mu_{12}} (1-z)^{\mu_{13}} G(z).
\end{equation}
Hence, one obtains
\begin{equation}
\begin{aligned}
&\left\{\frac{1}{\Delta_1+1}\partial_z^3
+2\left(\frac{1}{z}+\frac{1}{z-1}\right)\partial_z^2
+\left[\frac{\Delta_1-2\Delta_2}{z^2}+\frac{\Delta_1-2\Delta_3}{(z-1)^2}+\frac{3\Delta_1+2(\Delta_{23}+1)}{z(z-1)}\right]\partial_z
\right.\\
&\left. \quad -\Delta_1\left[\frac{2\Delta_2}{z^3}+\frac{2\Delta_3}{(z-1)^3}+(\Delta_1+\Delta_{23})
\left(\frac{1}{z^2}-\frac{1}{(z-1)^2}\right)\right]\right\} H(z)=0,
\end{aligned}
\end{equation}
where $\Delta_{23}=\Delta_2+\Delta_3-\Delta_4$.
This resembles the Belavin–Polyakov–Zamolodchikov (BPZ) equation \cite{Belavin:1984vu} for the 4-point correlator $\braket{\Delta(z)\mathcal{O}_1(0)\mathcal{O}_2(1)\mathcal{O}_3(\infty)}$.
Noting that the conformal weight for a $(1,3)$ vertex operator is given by 
\begin{equation}
    \Delta(1,3)=\alpha(Q-\alpha)= -\frac{1}{b}\left[b+\frac{1}{b}-\left(-\frac{1}{b}\right)\right]=-1-\frac{2}{b^2},
\end{equation}
one can also express this BPZ-type equation as follows:
\begin{equation}
\begin{aligned}
&\left\{-\frac{b^2}{2} \partial_z^3+2\left(\frac{1}{z}+\frac{1}{z-1}\right)\partial_z^2
+\left[\frac{\Delta_1-2\Delta_2}{z^2}+\frac{\Delta_1-2\Delta_3}{(z-1)^2}+\frac{3\Delta_1+2(\Delta_2+\Delta_3-\Delta_4+1)}{z(z-1)}\right]\partial_z
\right.\\
&\left. \quad +\left(\frac{2}{b^2}+1\right)\left[\frac{2\Delta_2}{z^3}+\frac{2\Delta_3}{(z-1)^3}
+(\Delta_1+\Delta_2+\Delta_3-\Delta_4)\left(\frac{1}{z^2}-\frac{1}{(z-1)^2}\right)\right] \right\} H(z)=0.
\end{aligned}
\end{equation}
This equation also applies to the primary field of Kac label $(3,1)$ which is degenerate at level 3, provided the substitution $b\to 1/b$ is made. 
\subsection{Regular blocks in the tricritical Ising model}
Consider the minimal model $\mathcal{M}(4,5)$ or the tricritical Ising model. In this model, the thermal field $\epsilon'$ obeys the same fusion rule as that of a Fibonacci anyon \cite{DiFrancesco:639405}:
\begin{equation}
\epsilon'\times\mathbf{1}=\epsilon', \quad
\epsilon'\times\epsilon'=\mathbf{1}+\epsilon',
\end{equation}
where $\mathbf{1}$ is the identity operator of conformal weight $0$.
The conformal weight of $\epsilon'$ is $\Delta=3/5$ while the central charge of the theory is $c=7/10$. $\epsilon'$ has Kac label $(r,s)=(1,3)$, so we can apply the general story in the last subsection. The correlator $\braket{\epsilon'(\infty)\epsilon'(1)\epsilon'(z)\epsilon'(0)}$ satisfies
\begin{equation}\label{eq:bpzfib}
\begin{aligned}
&\left[\frac{5}{8}\partial_z^3+2\left(\frac{1}{z}+\frac{1}{z-1}\right)\partial_z^2
-\left(\frac{3}{5z^2}+\frac{3}{5(z-1)^2}-\frac{5}{z(z-1)}\right)\partial_z
\right.\\
&\quad \left. -\frac{18}{25}\left(\frac{1}{z^3}+\frac{1}{(z-1)^3}+\frac{1}{z^2}-\frac{1}{(z-1)^2}\right)
\right]\braket{\epsilon'(\infty)\epsilon'(1)\epsilon'(z)\epsilon'(0)}=0.
\end{aligned}
\end{equation}
We stress that a correlator is a sesquilinear combination of conformal blocks. Therefore, the conformal blocks $\mathcal{B}_i(z)$ of $\braket{\epsilon'(\infty)\epsilon'(1)\epsilon'(z)\epsilon'(0)}$ also satisfy the above BPZ-type equation. There are three linearly-independent solutions to this equation, namely two of them are 
\begin{equation}\label{eq:rcbsol}
    \begin{aligned}
        \mathcal{B}_1(z)&=z^{-6/5}(z-1)^{-3/5}(z^2-z+1)\,_2F_1\left(\frac15,\frac45;\frac25;z\right),\\
        \mathcal{B}_2(z)&=z^{-3/5}(z-1)^{-3/5}(z^2-z+1) \,_2F_1\left(\frac45,\frac75;\frac85;z\right),
    \end{aligned}
\end{equation}
while the remaining solution is not in closed form. 

Given that the OPE of two primaries $\mathcal{O}_i(z_i)\times \mathcal{O}_j(z_j)\sim \sum_k(z_i-z_j)^{\varepsilon_k}\mathcal{O}_k(z_k)$ with $\varepsilon_k=\Delta_k-\Delta_i-\Delta_j$, the fusion rule of primaries in a block is determined by these exponents, which appear to be the leading contribution to the expansion $\mathcal{B}_i\sim z^{\varepsilon_i}(1+\mathcal{O}(z))$. Hence by expanding the solutions \eqref{eq:rcbsol} around $z=0$, we find that 
\begin{equation}
    \mathcal{B}_1(z)\sim z^{-6/5},\quad
    \mathcal{B}_2(z)\sim z^{-3/5},
\end{equation}
which correspond, respectively, to the fusion rules $\epsilon'\times \epsilon'\to \mathbf{1}$ and $\epsilon'\times \epsilon'\to \epsilon'$.

To investigate the monodromy behavior of these blocks, we will adopt an alternative approach by first representing $\epsilon$'s as vertex operators. This allows for the free field realization of conformal blocks, which turns the blocks $\mathcal{B}_i(z)$ into Coulomb gas integrals.

\begin{figure}[!ht]
\centering
\begin{tikzpicture}
\draw[black, thick] (-3,0)node[anchor=east]{$\epsilon'(\infty)$} -- (3,0)node[anchor=west]{$\epsilon'(0)$};
\draw[black, thick] (-1.5,1)node[anchor=south]{$\epsilon'(1)$} -- (-1.5,0);
\draw[black, thick] (1.5,1)node[anchor=south]{$\epsilon'(z)$} -- (1.5,0);
\node at (0,0.3) {$\mathcal{O}$};
\end{tikzpicture}
\caption{A regular 4-point $\epsilon'$ conformal block. $\mathcal{O}$ denotes the intermediate state formed by the fusion of two distinct $\epsilon'$ fields.}
\label{fig:fusionchannel}
\end{figure}
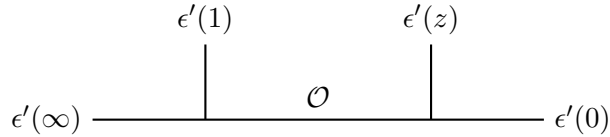

Taking the momentum of $\epsilon'$ field at $\infty$ to be $Q-\alpha$ (while the other three fields are of momenta $\alpha$), we are able to easily remove the background charge and minimize the number of screening charges needed for the charge neutrality condition. In our case, we have two screening charges $\mathcal{Q}_-$, each with momentum $\alpha_- =b^{-1}$ for this correlator. A general conformal block for $\braket{\epsilon'(\infty)\epsilon'(1)\epsilon'(z)\epsilon'(0)}$ is then given by
\begin{equation}\label{eq:rcb}
\begin{aligned}
   \mathcal{B}(z)&= z^{-2\alpha_0\alpha_1}(1-z)^{-2\alpha_1\alpha_2} \int_{\mathcal{C}_1} dw_1
   \int_{\mathcal{C}_2} dw_2\\
   &\quad \times 
   (w_1 w_2)^{-2\alpha_0 \alpha_-}
   [(w_1-z)(w_2-z)]^{-2\alpha_1 \alpha_-}
   [(w_1-1)(w_2-1)]^{-2\alpha_2 \alpha_-}
   (w_1-w_2)^{-2\alpha_-^2}\\
   &= \left[z(1-z)\right]^{-\frac{2}{b^2}} \int_{\mathcal{C}_1} dw_1
   \int_{\mathcal{C}_2} dw_2\,
   (w_1 w_2)^{\frac{2}{b^2}}
   [(w_1-z)(w_2-z)]^{\frac{2}{b^2}}
   [(w_1-1)(w_2-1)]^{\frac{2}{b^2}}
   (w_1-w_2)^{-\frac{2}{b^2}}.
\end{aligned}
\end{equation}
This 4-point $\epsilon'$ conformal block can be represented as a comb diagram in Figure \ref{fig:fusionchannel}. 

\subsection{Analytic continuation}
There are various ways of positioning the integration contours for the block \eqref{eq:rcb}. It would be instructive to use contours that are open paths between the singularities. These contours are homotopic to the 1-dimensional Lefschetz thimbles, as mentioned in the previous section. Alternatively, these integrals can be obtained by shrinking closed contours surrounding the singularities. One has the choice of integrating each variable $w_1$ and $w_2$ along one of the four contours:
$-\infty\to 0, 0\to z, z\to 1, 1\to \infty$.
Moreover, the paths of $w_1$ and $w_2$ are not allowed to intersect. One would also need to specify which path lies on top of the other.

\begin{figure}[!ht]\centering
\tikzset{every picture/.style={line width=0.75pt}}    
\begin{tikzpicture}[x=0.75pt,y=0.75pt,yscale=-1,xscale=1]
\draw    (136.01,74) -- (180.31,74) ;
\draw    (136.01,148.49) -- (180.31,148.49) ;
\draw  [color={rgb, 255:red, 208; green, 2; blue, 27 }  ,draw opacity=1 ] (122,81.5) .. controls (122,70.45) and (137.67,61.5) .. (157,61.5) .. controls (176.33,61.5) and (192,70.45) .. (192,81.5) .. controls (192,92.55) and (176.33,101.5) .. (157,101.5) .. controls (137.67,101.5) and (122,92.55) .. (122,81.5) -- cycle ;
\draw  [color={rgb, 255:red, 126; green, 211; blue, 33 }  ,draw opacity=1 ] (123,156.5) .. controls (123,145.45) and (138.67,136.5) .. (158,136.5) .. controls (177.33,136.5) and (193,145.45) .. (193,156.5) .. controls (193,167.55) and (177.33,176.5) .. (158,176.5) .. controls (138.67,176.5) and (123,167.55) .. (123,156.5) -- cycle ;
\draw    (274.01,71) -- (318.31,71) ;
\draw    (274.01,145.49) -- (318.31,145.49) ;
\draw  [color={rgb, 255:red, 74; green, 144; blue, 226 }  ,draw opacity=1 ] (260,78.5) .. controls (260,67.45) and (275.67,58.5) .. (295,58.5) .. controls (314.33,58.5) and (330,67.45) .. (330,78.5) .. controls (330,89.55) and (314.33,98.5) .. (295,98.5) .. controls (275.67,98.5) and (260,89.55) .. (260,78.5) -- cycle ;
\draw  [color={rgb, 255:red, 126; green, 211; blue, 33 }  ,draw opacity=1 ] (261,153.5) .. controls (261,142.45) and (276.67,133.5) .. (296,133.5) .. controls (315.33,133.5) and (331,142.45) .. (331,153.5) .. controls (331,164.55) and (315.33,173.5) .. (296,173.5) .. controls (276.67,173.5) and (261,164.55) .. (261,153.5) -- cycle ;
\draw    (373.51,69) -- (417.81,69) ;
\draw    (373.51,143.49) -- (417.81,143.49) ;
\draw  [color={rgb, 255:red, 208; green, 2; blue, 27 }  ,draw opacity=1 ] (360.5,76.5) .. controls (360.5,65.45) and (376.17,56.5) .. (395.5,56.5) .. controls (414.83,56.5) and (430.5,65.45) .. (430.5,76.5) .. controls (430.5,87.55) and (414.83,96.5) .. (395.5,96.5) .. controls (376.17,96.5) and (360.5,87.55) .. (360.5,76.5) -- cycle ;
\draw  [color={rgb, 255:red, 245; green, 166; blue, 35 }  ,draw opacity=1 ] (361.5,151.5) .. controls (361.5,140.45) and (377.17,131.5) .. (396.5,131.5) .. controls (415.83,131.5) and (431.5,140.45) .. (431.5,151.5) .. controls (431.5,162.55) and (415.83,171.5) .. (396.5,171.5) .. controls (377.17,171.5) and (361.5,162.55) .. (361.5,151.5) -- cycle ;
\draw  [color={rgb, 255:red, 74; green, 144; blue, 226 }  ,draw opacity=1 ] (115.44,81.5) .. controls (115.44,68.38) and (134.05,57.75) .. (157,57.75) .. controls (179.95,57.75) and (198.56,68.38) .. (198.56,81.5) .. controls (198.56,94.62) and (179.95,105.25) .. (157,105.25) .. controls (134.05,105.25) and (115.44,94.62) .. (115.44,81.5) -- cycle ;
\draw  [color={rgb, 255:red, 245; green, 166; blue, 35 }  ,draw opacity=1 ] (116.44,156.5) .. controls (116.44,143.38) and (135.05,132.75) .. (158,132.75) .. controls (180.95,132.75) and (199.56,143.38) .. (199.56,156.5) .. controls (199.56,169.62) and (180.95,180.25) .. (158,180.25) .. controls (135.05,180.25) and (116.44,169.62) .. (116.44,156.5) -- cycle ;

\draw (123,15) node [anchor=north west][inner sep=0.75pt]   [align=left] {\textbf{$w_1$ contour}};
\draw (307,14) node [anchor=north west][inner sep=0.75pt]   [align=left] {\textbf{$w_2$ contour}};
\draw (130.01,76.36) node [anchor=north west][inner sep=0.75pt]    {$0$};
\draw (173.71,76.36) node [anchor=north west][inner sep=0.75pt]    {$z$};
\draw (130.37,151.27) node [anchor=north west][inner sep=0.75pt]    {$1$};
\draw (169.78,151.27) node [anchor=north west][inner sep=0.75pt]    {$\infty $};
\draw (268.01,73.36) node [anchor=north west][inner sep=0.75pt]    {$0$};
\draw (304.71,73.36) node [anchor=north west][inner sep=0.75pt]    {$w_{1}$};
\draw (268.37,148.27) node [anchor=north west][inner sep=0.75pt]    {$0$};
\draw (312.78,148.27) node [anchor=north west][inner sep=0.75pt]    {$1$};
\draw (367.51,71.36) node [anchor=north west][inner sep=0.75pt]    {$1$};
\draw (411.21,71.36) node [anchor=north west][inner sep=0.75pt]    {$\infty $};
\draw (367.87,146.27) node [anchor=north west][inner sep=0.75pt]    {$w_{1}$};
\draw (407.28,146.27) node [anchor=north west][inner sep=0.75pt]    {$\infty $};
\draw (386,37.9) node [anchor=north west][inner sep=0.75pt]  [color={rgb, 255:red, 208; green, 2; blue, 27 }  ,opacity=1 ]  {$\mathcal{B}_{1}$};
\draw (285,39.9) node [anchor=north west][inner sep=0.75pt]  [color={rgb, 255:red, 74; green, 144; blue, 226 }  ,opacity=1 ]  {$\mathcal{B}_{2}$};
\draw (388,113.4) node [anchor=north west][inner sep=0.75pt]  [color={rgb, 255:red, 245; green, 166; blue, 35 }  ,opacity=1 ]  {$\mathcal{B}_{4}$};
\draw (286,114.9) node [anchor=north west][inner sep=0.75pt]  [color={rgb, 255:red, 126; green, 211; blue, 33 }  ,opacity=1 ]  {$\mathcal{B}_{3}$};
\end{tikzpicture}
    \caption{Choices of domains of integration for the 4-point conformal blocks of $\epsilon'$ fields.}
    \label{fig:dintcb}
\end{figure}
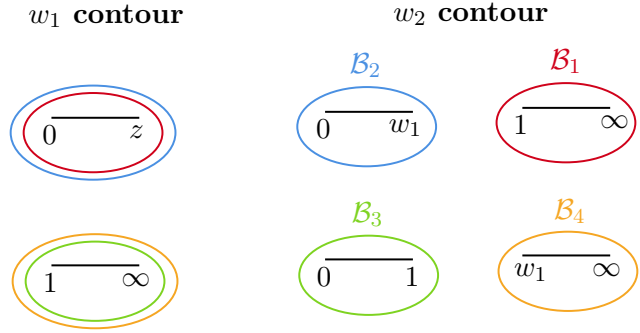

Here we focus on the integration cycles $0\to z$ and $1\to \infty$. These are the $s$-channel basis that correspond to the conformal blocks with abelian monodromy around $z=0$ \cite{Belavin:2016qaa}. (The cycles $-\infty\to 0$ and $z\to 1$ correspond to the $t$-channel basis.)
These blocks can be expanded as formal power series $\mathcal{B}_i(z)=z^{\nu_i}\sum_{n= 0}^\infty a_n^iz^n$, where $z^{\nu_i}$ is the leading contribution at small $z$ expansion. 
We have 4 possible choices of contours, as illustrated in Figure \ref{fig:dintcb}. 

Consider one of the conformal blocks for instance, 
\begin{equation}
    \mathcal{B}_1(z)=
    \left[z(1-z)\right]^{-\frac{2}{b^2}} \int_0^z dw_1
   \int_1^\infty dw_2\,
   (w_1 w_2)^{\frac{2}{b^2}}
   [(w_1-z)(w_2-z)]^{\frac{2}{b^2}}
   [(w_1-1)(w_2-1)]^{\frac{2}{b^2}}
   (w_1-w_2)^{-\frac{2}{b^2}}.
\end{equation}
With a simple change of variable $w_1\to z w_1$, we have 
\begin{equation}
\begin{aligned}
    \mathcal{B}_1(z) &=
    z^{1+\frac{4}{b^2}}[z(1-z)]^{-\frac{2}{b^2}} \int_0^1 dw_1
   \int_1^\infty dw_2\\
   &\quad\times (w_1 w_2)^{\frac{2}{b^2}}
   [(w_1-1)(w_2-z)]^{\frac{2}{b^2}}
   [(z w_1-1)(w_2-1)]^{\frac{2}{b^2}}
   (z w_1-w_2)^{-\frac{2}{b^2}}.
\end{aligned}
\end{equation}
We can express this in exponential form as follows, 
\begin{equation}
   \mathcal{B}_1(z)
   = z \int_0^1 dw_1
   \int_1^\infty dw_2\, \exp{\left(\frac{1}{b^2}f(z,w_1,w_2)\right)},
\end{equation}
where $f(z,w_1,w_2)$ is given by
\begin{equation}\label{eq:yyr}
\begin{aligned}
    f(z,w_1,w_2)&=2\left[\log z-\log(1-z)\right]-2\log(zw_1-w_2)\\
    &\quad +2\left[\log w_1+\log w_2+\log(w_1-1)+\log (w_2-z)+\log(zw_1-1)+\log(w_2-1)\right].
\end{aligned}
\end{equation}
Given that an exponent has the property that for an integer $n\in \mathbb{Z}$, $\exp(x)= \exp(x+2 \pi i n)$, each time we circle the contour round a singularity, we will pick up a phase $\exp(2 \pi i)$. Depending on the orientation of each loop, we will get an overall phase in terms of the parameter $q=\exp{\left(-\frac{2\pi i}{b^2}\right)}$.
Thus we stress that the integrand we have is multi-valued, a property which is at play in the evaluation below. 

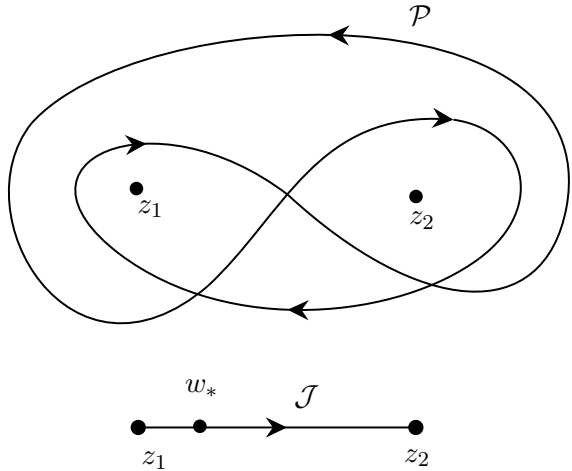
\begin{figure}[!ht]\centering
\tikzset{every picture/.style={line width=0.75pt}} 
\begin{tikzpicture}[x=0.75pt,y=0.75pt,yscale=-1,xscale=1]
\draw  [fill={rgb, 255:red, 0; green, 0; blue, 0 }  ,fill opacity=1 ] (126,130.42) .. controls (126,128.81) and (127.31,127.5) .. (128.92,127.5) .. controls (130.53,127.5) and (131.83,128.81) .. (131.83,130.42) .. controls (131.83,132.03) and (130.53,133.33) .. (128.92,133.33) .. controls (127.31,133.33) and (126,132.03) .. (126,130.42) -- cycle ;
\draw    (120.53,163.58) .. controls (176.53,205.58) and (261.83,195.67) .. (301.83,165.67) ;
\draw [shift={(205.12,191.58)}, rotate = 1.15] [fill={rgb, 255:red, 0; green, 0; blue, 0 }  ][line width=0.08]  [draw opacity=0] (10.72,-5.15) -- (0,0) -- (10.72,5.15) -- (7.12,0) -- cycle    ;
\draw    (120.53,163.58) .. controls (58.53,116.58) and (137.53,81.58) .. (205.3,133.58) ;
\draw [shift={(134.01,107.77)}, rotate = 176.22] [fill={rgb, 255:red, 0; green, 0; blue, 0 }  ][line width=0.08]  [draw opacity=0] (10.72,-5.15) -- (0,0) -- (10.72,5.15) -- (7.12,0) -- cycle    ;
\draw    (205.3,133.58) .. controls (271.3,194.58) and (334.53,200.58) .. (345.83,139.67) ;
\draw    (75.83,97.67) .. controls (130.83,35.67) and (365.83,27.67) .. (345.83,139.67) ;
\draw [shift={(225.58,52.84)}, rotate = 1.69] [fill={rgb, 255:red, 0; green, 0; blue, 0 }  ][line width=0.08]  [draw opacity=0] (10.72,-5.15) -- (0,0) -- (10.72,5.15) -- (7.12,0) -- cycle    ;
\draw    (164.83,179.67) .. controls (95.83,238.67) and (39.83,143.67) .. (75.83,97.67) ;
\draw    (164.83,179.67) .. controls (203.44,144.03) and (219.51,90.74) .. (286.78,95.5) ;
\draw [shift={(288.83,95.67)}, rotate = 184.97] [fill={rgb, 255:red, 0; green, 0; blue, 0 }  ][line width=0.08]  [draw opacity=0] (10.72,-5.15) -- (0,0) -- (10.72,5.15) -- (7.12,0) -- cycle    ;
\draw    (288.83,95.67) .. controls (324.83,100.67) and (336.83,139.67) .. (301.83,165.67) ;
\draw    (129.83,251) -- (269.83,251) ;
\draw [shift={(269.83,251)}, rotate = 0] [color={rgb, 255:red, 0; green, 0; blue, 0 }  ][fill={rgb, 255:red, 0; green, 0; blue, 0 }  ][line width=0.75]      (0, 0) circle [x radius= 3.35, y radius= 3.35]   ;
\draw [shift={(204.83,251)}, rotate = 180] [fill={rgb, 255:red, 0; green, 0; blue, 0 }  ][line width=0.08]  [draw opacity=0] (10.72,-5.15) -- (0,0) -- (10.72,5.15) -- (7.12,0) -- cycle    ;
\draw [shift={(129.83,251)}, rotate = 0] [color={rgb, 255:red, 0; green, 0; blue, 0 }  ][fill={rgb, 255:red, 0; green, 0; blue, 0 }  ][line width=0.75]      (0, 0) circle [x radius= 3.35, y radius= 3.35]   ;
\draw  [fill={rgb, 255:red, 0; green, 0; blue, 0 }  ,fill opacity=1 ] (267,134.42) .. controls (267,132.81) and (268.31,131.5) .. (269.92,131.5) .. controls (271.53,131.5) and (272.83,132.81) .. (272.83,134.42) .. controls (272.83,136.03) and (271.53,137.33) .. (269.92,137.33) .. controls (268.31,137.33) and (267,136.03) .. (267,134.42) -- cycle ;
\draw  [fill={rgb, 255:red, 0; green, 0; blue, 0 }  ,fill opacity=1 ] (158,250.42) .. controls (158,248.81) and (159.31,247.5) .. (160.92,247.5) .. controls (162.53,247.5) and (163.83,248.81) .. (163.83,250.42) .. controls (163.83,252.03) and (162.53,253.33) .. (160.92,253.33) .. controls (159.31,253.33) and (158,252.03) .. (158,250.42) -- cycle ;

\draw (128,134.82) node [anchor=north west][inner sep=0.75pt]    {$z_{1}$};
\draw (264.92,140.73) node [anchor=north west][inner sep=0.75pt]    {$z_{2}$};
\draw (130,261.9) node [anchor=north west][inner sep=0.75pt]    {$z_{1}$};
\draw (262.5,260.4) node [anchor=north west][inner sep=0.75pt]    {$z_{2}$};
\draw (265,36.4) node [anchor=north west][inner sep=0.75pt]    {$\mathcal{P}$};
\draw (207,228.4) node [anchor=north west][inner sep=0.75pt]    {$\mathcal{J}$};
\draw (152,225.4) node [anchor=north west][inner sep=0.75pt]    {$w_{*}$};
\end{tikzpicture}
    \caption{Pochhammer contour $\mathcal{P}$ (top) and its equivalent shrunken contour (bottom), which is homotopic to a 1-dimensional Lefschetz thimble $\mathcal{J}$. $w_*$ is the critical point of the Yang-Yang function $\mathcal{W}$.}
    \label{fig:pochh}
\end{figure}

One well-defined closed integration contour is the Pochhammer contour $\mathcal{P}$ (Figure \ref{fig:pochh}), which loops twice around the singularities. This contour is equivalent to the thimble $\mathcal{J}$ (connecting $z=0$ with $z=1$) up to a prefactor in the form of a finite Laurent polynomial $L(q)$. This can be shown by fixing $w_2\in(1,\infty)$ and moving $w_1$. By adopting the contour $\mathcal{P}$, we have analytically continued the integration over $w_1$ from the real line to the complex plane, and obtain the result as follows: 
\begin{equation}
\begin{aligned}
    \mathcal{B}_1(z)& =
    z^{1+\frac{4}{b^2}}[z(1-z)]^{-\frac{2}{b^2}} \int_{\mathcal{P}} dw_1
   \int_1^\infty dw_2\\
   &\quad\times (w_1 w_2)^{\frac{2}{b^2}}
   [(w_1-1)(w_2-z)]^{\frac{2}{b^2}}
   [(z w_1-1)(w_2-1)]^{\frac{2}{b^2}}
   (z w_1-w_2)^{-\frac{2}{b^2}}\\
   &= (1-q^{-2})^2 z^{1+\frac{4}{b^2}}[z(1-z)]^{-\frac{2}{b^2}} \int_0^1 dw_1
   \int_1^\infty dw_2\\
   &\quad\times (w_1 w_2)^{\frac{2}{b^2}}
   [(w_1-1)(w_2-z)]^{\frac{2}{b^2}}
   [(z w_1-1)(w_2-1)]^{\frac{2}{b^2}}
   (z w_1-w_2)^{-\frac{2}{b^2}}.
   \end{aligned}
\end{equation}

Since our differential equation \eqref{eq:bpzfib} is of order 3, we expect three linearly-independent solutions corresponding to three distinct conformal blocks.
The Coulomb gas integrals for these blocks provide a transparent way to construct the monodromy matrices $\mathcal{M}_0$ and $\mathcal{M}_1$ around $z=0$ and $z=1$, which generate the monodromy group. We find that $\mathcal{M}_0$ is a diagonal matrix,
\begin{equation}
    \mathcal{M}_0=\operatorname{diag}\left(q^{4}, q^{4}, 1 \right),
\end{equation}
where this follows from the monodromy around $z=0$ for the blocks $\mathcal{B}_1$, $\mathcal{B}_2$, and $\mathcal{B}_3$.
We have chosen one of the blocks from $\mathcal{B}_3$ and $\mathcal{B}_4$ since they behave identically for small $z$.
On the other hand, $\mathcal{M}_1$ is not diagonal. However we can relate it to the diagonalized matrix $\widetilde{\mathcal{M}}_1=\operatorname{diag}\left(q^{4}, q^{4},1 \right)$, which is obtained from reexpressing $\mathcal{B}_i(z)$ in terms of $\mathcal{B}_i(1-z).$
Then we apply the basis transformation 
\begin{equation}
    \mathcal{M}_1= U^{-1} \widetilde{\mathcal{M}}_1 U,
\end{equation}
in which the $3\times 3$ matrix $U$ relates $\mathcal{B}_i(z)$ to other blocks $\mathcal{B}_j(z), j\neq i$ and the blocks $\widetilde{\mathcal{B}}(z)$ whose contours run from $-\infty\to 0$ and $z\to 1$:
\begin{equation}
    \widetilde{\mathcal{B}}_i= U_{ij} \mathcal{B}_j.
\end{equation}
The complete relations between the blocks can be obtained via the application of the Cauchy integral theorem, which deforms a certain contour into a linear combination of complementary contours. For instance, the contour of $w_1$: $[0,z]$ can be deformed into $(-\infty,0]\cup [z,\infty)$. For detailed analysis of the basis transformation, we refer the interested reader to \cite{DiFrancesco:639405, Belavin:2016qaa}.

\section{Irregular conformal blocks}\label{sec:irreg}
In this section we would like to extend the study of conformal blocks to a more general case where one of the fields is an exotic state, commonly known as the \emph{irregular state} \cite{Gaiotto:2009ma} or \emph{wild state} \cite{Bonelli:2011aa}.
To this end, we consider a generalised vertex operator under the Coulomb gas formalism:
\begin{equation}
    I_{\alpha,\beta}(z)=:\exp(2\alpha \phi(z))\exp(\beta \partial \phi(z)):.
\end{equation}
When $\beta=0$, this operator $I$ reduces to an ordinary vertex operator. $\alpha$ is the $U(1)$ charge and we should also impose the charge neutrality condition on correlators involving $I_{\alpha,\beta}$. The correlator of this field with ordinary vertex operators in the free theory reads 
\begin{equation}
\label{eq:irregularcorre}
     \left\langle\prod_{i=1}^n V_{\alpha_i}(z_i)I_{\alpha,\beta}(z)\right\rangle=
     \begin{cases}
        \prod_{i<j}(z_i-z_j)^{-2\alpha_i \alpha_j}(z_i-z)^{-2\alpha_i\alpha}\exp\left(-\frac{\alpha_i\beta}{z_i-z}\right) & \text{if } \sum_{i=1}^n \alpha_i+\alpha=Q,\\ 
        0 & \text{otherwise. }
    \end{cases}
\end{equation}
If we want to represent the conformal blocks in minimal models (say, the tricritical Ising model in the earlier sections), we need also to use the screening charges and contour integrals. One can also utilize the symmetry of Kac labels \eqref{eq:weylpotts} as in the regular case.

\subsection{Irregular states}
Here we give a brief introduction to irregular states, which correspond to the $I_{\alpha,\beta}$ operator mentioned above.

The stress tensor $T$ in the Liouville field theory is given by
\begin{equation}
    T(z)=-:\partial_z \phi(z) \partial_z \phi(z): +Q\partial^2_z \phi(z),
\end{equation}
with $Q=b+1/b$. We can compute the OPE of $T$ and $I_{\alpha,\beta}$ as in \cite{haghighat2023flat}:
\begin{equation}
    \begin{aligned}
        T(z)I_{\alpha,\beta}(w)&\sim \left(-\frac{\beta^4}{4}\frac{1}{(z-w)^4}+\frac{\beta(Q-\alpha)}{(z-w)^3}+\frac{\alpha(Q-\alpha)}{(z-w)^2}\right)I_{\alpha,\beta}(w)\\\
        &\quad+\left(\frac{\beta}{(z-w)^2}+\frac{\alpha+\beta}{(z-w)}\right)\partial\phi(w)I_{\alpha,\beta}(w).
    \end{aligned}
\end{equation}
We define $\ket{I}$ to be the state corresponding to the $I_{\alpha,\beta}$ operator acting on the vacuum:
\begin{equation}
    \ket{I}=I_{\alpha,\beta}(0)\ket{0}.
\end{equation}
Using the OPE of $T$ and $I_{\alpha,\beta}$ with the definition of the Virasoro generators 
\begin{equation}
    L_n=\oint\frac{dz}{2\pi i}z^{n+1}T(z),
\end{equation}
one obtains the actions of $L_n$ on $\ket{I}$:
\begin{equation}\label{eq:vgir}
    \begin{aligned}
& L_0 \ket{I} =\Delta_\alpha+\beta \partial_{\beta}\ket{I} \\
& L_{1}\ket{I} =\beta (Q-\alpha)\ket{I} \\
& L_{2}\ket{I}=-\frac{\beta^2}{4}\ket{I} \\
& L_{n}\ket{I} =0, \quad n>2.
\end{aligned}
\end{equation}

These properties actually define an irregular state of rank $1$. In general, an irregular state $\ket{I_\kappa}$ of rank $\kappa\in \mathbb{Z}_{\geq 0}$ is a simultaneous eigenstate of the positive Virasoro generators $L_{i}$ for $\kappa\leq i\leq 2\kappa$, but is annihilated by $L_{i}$ for $i>2\kappa$. Denoting the set of eigenvalues as $\Lambda=\{\Lambda_i\}_{i=\kappa,\cdots,2\kappa}$ with $\Lambda_i\neq 0$, a rank $\kappa$ irregular state satisfies the eigenvalue equations:
\begin{equation}
\begin{aligned}
L_i\ket{I_\kappa}&=\Lambda_i\ket{I_\kappa}, \quad i=\kappa,\dots,2\kappa,\\
        L_i\ket{I_\kappa}&=0, \quad i>2\kappa.
\end{aligned}    
\end{equation}
One can construct an irregular state via the collision limit of two or more primary states, as outlined by Gaiotto and Teschner \cite{Gaiotto:2012sf}.

\subsection{The tricritical Ising model again}
Back to the tricritical Ising model, we would like to study the correlators with an irregular field,
\begin{equation}
\label{eq:irregular}
\langle \epsilon'(z_1)\epsilon'(z_2)\epsilon'(z_3)I_{\alpha,\beta}(z_4)\rangle.
\end{equation}
There are many possible ways to express this correlator as a Coulomb gas integral, because we can choose freely the number of screening charges and the value of $\alpha$. For our purpose, we adopt one screening current $V_{\alpha_-}(w)$, $\alpha=0$ and take the momentum of $\epsilon'(z_3)$ to be $Q-\alpha_3$. In other words, we have the four fields as follows:
\begin{equation}
\begin{aligned}
        \epsilon'(z_{1,2})&=\, :\exp\left(-\frac2b\phi(z_{1,2})\right):,\\
        \epsilon'(z_{3})&=\, :\exp\left(2\left(b+\frac2b\right)\phi(z_{3})\right):,\\
        I_{\beta}(z_4)&=\, :\exp(\beta \partial\phi(z_4)):.
\end{aligned}
\end{equation}
One can verify that this combination indeed satisfies the neutrality condition.

Then, we would like to send the coordinate $z_4$ to infinity. Naively, this would reduce the correlator to a regular one, because \eqref{eq:irregularcorre} tells us there are factors of $\exp{\frac{2\beta/b}{z_i-z_4}}$.
However, instead, we can let $\beta$ vary with $z_4$, $\beta=\frac{z_4^2}{b}\Lambda$, where $\Lambda$ is an auxiliary constant. Then
\begin{equation}
\begin{aligned}
      \exp\left(\frac{2\beta/b}{z_i-z_4}\right)&\sim\exp\left[-\frac{2\Lambda z_4}{b^2}\left(1+\frac{z_i}{z_4}+O(z_4^2)\right)\right] \\
      &\propto \exp\left(-\frac{2\Lambda}{b^2}z_i\right),
\end{aligned}
\end{equation}
where in the last line we omit the infinite proportionality factor $\exp\left(-\frac{2z_4}{b^2}\Lambda\right)$.
We denote this limiting state as $I_\Lambda(\infty)$ and \eqref{eq:irregular} finally becomes
\begin{equation}\label{eq:irregcf}
\begin{aligned}
\langle \epsilon'(z_1)\epsilon'(z_2)\epsilon'(z_3)I_\Lambda(\infty)\rangle
&=(z_1-z_2)^{8/5}(z_1-z_3)^{-6/5}(z_2-z_3)^{-6/5}e^{(-\frac45z_1-\frac45z_2+\frac35z_3)\Lambda}\\
&\quad \times\int dw  (z_1-w)^{-8/5}(z_2-w)^{-8/5}(z_3-w)^{6/5}e^{\frac45  \Lambda w}.
\end{aligned}
\end{equation}
Here we take the integration contour to be thimbles. For simplicity, we will henceforth denote $\langle \epsilon'(z_1)\epsilon'(z_2)\epsilon'(z_3)I_\Lambda(\infty)\rangle$ as $F(z_1,z_2,z_3,\Lambda)$.

\subsection{Irregular block equations}
We would like to show that $F(z_1,z_2,z_3,\Lambda)$ satisfies the following $3$ differential equations:
\begin{equation}\label{eq:gci1}
\left(\partial_{z_1}+\partial_{z_2}+\partial_{z_3}+\frac{Q\Lambda}{b}\right)F(z_1,z_2,z_3,\Lambda)=0,
\end{equation}
\begin{equation}\label{eq:gci2}
\left(z_1\partial_{z_1}+z_2\partial_{z_2}+z_3\partial_{z_3}+3\Delta-\Lambda\partial_\Lambda\right)F(z_1,z_2,z_3,\Lambda)=0,
\end{equation}
\begin{equation}\label{eq:nullvec}
\begin{aligned}
    D F(z_1,z_2,z_3,\Lambda) &=\left[\frac{2\Delta}{(z_3-z_1)^3} +\frac{2\Delta}{(z_2-z_1)^3}-\frac{1}{(z_2-z_1)^2}\partial_{z_2}-\frac{1}{(z_3-z_1)^2}\partial_{z_3}\right.\\
    &\quad-\frac{2}{\Delta}\left(
    \frac{\Delta}{(z_3-z_1)^2}+\frac{\Delta}{(z_2-z_1)^2}-\frac{1}{z_2-z_1}\partial_{z_2}-\frac{1}{z_3-z_1}\partial_{z_3}-\frac{\Lambda^2}{4b^2}\right)\partial_{z_1}\\
    &\quad \left.+\frac{1}{\Delta(\Delta+1)}\partial_{z_1}^3\right]F(z_1,z_2,z_3,\Lambda)=0.
\end{aligned}
\end{equation}
with $b=i\frac{\sqrt{5}}{2}$, $Q=b+1/b$, $\Delta=3/5$. They correspond to $L_{-1}$ global conformal invariance, $L_0$ global conformal invariance and level $3$ null vector equation, respectively. They can be derived from the OPEs of $T$ with the regular and irregular operators, as well as the residue theorem \cite{haghighat2023flat}. Here we want to explicitly verify that \eqref{eq:irregcf} is a solution of these three equations. Let us decompose $F(z_1,z_2,z_3,\Lambda)$ into the overall factor part and the integral part:
\begin{equation}
    F=f(z_1,z_2,z_3)\int \phi(z_1,z_2,z_3,w)\,dw.
\end{equation}
Then the first equation \eqref{eq:gci1} becomes
\begin{equation}
    \begin{aligned}
\left(\partial_{z_1}+\partial_{z_2}+\partial_{z_3}+\frac{Q\Lambda}{b}\right)F=-f\int\partial_w\phi\, dw=0. 
    \end{aligned}
\end{equation}
This vanishes because it is an integration over a total derivative, and we take the integration contour to be thimbles, where $\phi$ goes to zero at the endpoints.
Similarly, for the second equation \eqref{eq:gci2},
\begin{equation}
    \begin{aligned}
\left(z_1\partial_{z_1}+z_2\partial_{z_2}+z_3\partial_{z_3}-\Lambda\partial_\Lambda\right)F =
-f\int\partial_w(w\phi)\,dw=0.
    \end{aligned}
\end{equation}

The third equation is of order 3, which is more involved. Our strategy is to reduce the higher order differential operators into first order ones. Firstly, we can directly verify that the overall factor $f$ satisfies \eqref{eq:nullvec} as a function itself,
\begin{equation}
    Df=0.
\end{equation}
Secondly, we have a list of derivative relations:
\begin{equation}
    \begin{aligned}
\partial_{z_1}f=\frac25\left(\frac{4}{z_1-z_2}-\frac{3}{z_1-z_3}-2\Lambda\right)f.
    \end{aligned}
\end{equation}
Considering these equations, one can reduce \eqref{eq:nullvec} into an equation purely of $\int\phi \,dw$. Then we can reduce the order using integration by parts:
\begin{equation}
\begin{aligned}
    \partial_{z_1}^3\int \phi dw&=\left[\frac{8}{5}\frac{1}{(z_1-z_2)^2}(\partial_{z_1}-\partial_{z_2})+\frac{8}{5}\frac{1}{z_2-z_1}(\partial_{z_1}^2-\partial_{z_1}\partial_{z_2})\right.\\
      &\quad -\frac{6}{5}\frac{1}{(z_1-z_3)^2}\partial_{z_1}-\frac{8}{5}\frac{1}{(z_1-z_3)^2}\partial_{z_3}-\frac{6}{5}\frac{1}{z_3-z_1}\partial_{z_1}^2\\
      &\left.\quad -\frac{8}{5}\frac{1}{z_3-z_1}\partial_{z_1}\partial_{z_3}+\frac{4}{5}\Lambda \partial_{z_1}^2 \right]\int \phi dw.
\end{aligned}
\end{equation}
Also, we have
\begin{equation}
    \partial_{z_1}^2\int \phi dw=\left[-\partial_{z_1}\partial_{z_2}-\partial_{z_1}\partial_{z_3}+\frac{4}{5}\Lambda \partial_{z_1}\right]\int \phi dw.
\end{equation}
Then we can split the mixed derivatives into first-order operators:
\begin{equation}
    \begin{aligned}
\partial_{z_1}\partial_{z_2}\int \phi dw&=\frac{8}{5}\frac{1}{z_1-z_2}(\partial_{z_1}-\partial_{z_2})\int \phi dw,\\
\partial_{z_1}\partial_{z_3}\int \phi dw&=
\left(\frac{6}{5}\frac{1}{z_3-z_1}\partial_{z_1}+\frac{8}{5}\frac{1}{z_3-z_1}\right)\int \phi dw.
    \end{aligned}
\end{equation}
These equations suffice to reduce $DF=0$ into a first-order equation of $\int\phi dw$, and one can find that terms cancel each other. So we have verified indeed $DF=0$.

\subsection{Semiclassical limit}
Following \cite{Bonelli_2022}, let us consider the semiclassical limit of the irregular $\epsilon'$ conformal blocks. The semiclassical limit of Liouville theory corresponds to taking a double-scaling limit known as the Nekrasov-Shatashvili (NS) limit in the AGT dual gauge theory. 

We first combine the three equations \eqref{eq:gci1}, \eqref{eq:gci2} and \eqref{eq:nullvec} into one single ODE, just as we did in the regular case. The resulting differential equation is 
\begin{equation}\label{eq:irrbpz}
\begin{aligned}
&\left\{-\frac{b^2}{2}\partial_z^3
        +2\left( \frac{2z-1}{z(z-1)}\partial_{z}^2
         -\frac{1}{z(z-1)}\Lambda\partial_{\Lambda}\partial_z\right)
         -\left(\frac{2}{b^2}+1\right)\frac{1-2z}{z^2(z-1)^2}\Lambda\partial_{\Lambda}\right.\\
&\quad +2\left(\frac{3\Delta_1}{z(z-1)}
        -\frac{\Delta_2}{z^2}
        -\frac{\Delta_3}{(z-1)^2}
        -\Delta_1 \frac{3z(z-1)+1}{z^2(z-1)^2}
        -\frac{\mu\Lambda}{z}
        +\frac{\Lambda^2}{4b^2}\right)\partial_z\\
&\quad \left.+\left(\frac{2}{b^2}+1\right) \left( 
         3\Delta_1\frac{1-2z}{z^2(z-1)^2}
        +\frac{2\Delta_2}{z^3}
        +\frac{2\Delta_3}{(z-1)^3}
        +\frac{\mu\Lambda}{z^2}\right)
\right\}F(z,\Lambda)=0
\end{aligned}
\end{equation}
with $\mu=Q/b$. $\Delta_{1,2,3}$ are the conformal weights of $\epsilon'$ fields at $z,0$ and $1$ respectively; we intend to keep these notations for clarity.
(The derivation of the level 3 BPZ-type equation for a general correlator of three regular states and one irregular state can be found in Appendix \ref{app:irreg}.) 

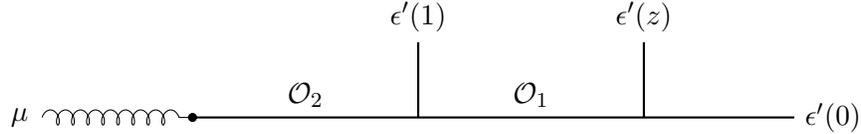
\begin{figure}[!ht]
\centering
\begin{tikzpicture}
\draw[black, thick] (-3,0) -- (5,0)node[anchor=west]{$\epsilon'(0)$};
\draw[black, thick] (3,1)node[anchor=south]{$\epsilon'(z)$} -- (3,0);
\draw[black, thick] (0,1)node[anchor=south]{$\epsilon'(1)$} -- (0,0);
\node at (1.5,0.3) {$\mathcal{O}_1$};
\node at (-1.5,0.3) {$\mathcal{O}_2$};
\node at (-5.3,0){$\mu$};
\filldraw [black] (-3,0) circle (1.5pt);
\path [draw, coil it]
    (-5,0)--(-3,0);
\end{tikzpicture}
\caption{Rank 1 irregular 4-point conformal block. $\mathcal{O}_1$ and $\mathcal{O}_2$ denote the intermediate states as the fusion products of two $\epsilon'$ fields. The curly line denotes the rank 1 irregular state and the black dot the projection onto the primary state $\mathcal{O}_2$.}
\label{fig:4pticb}
\end{figure}

The semiclassical limit amounts to taking a large central charge $c\to\infty$. More precisely, we take $b\to \infty$ and $\alpha_i\to \infty,\Lambda\to\infty$, while keeping $a_i=\alpha_i/b$ and $L=\Lambda/{b^2}$ finite. 
Schematically, the conformal blocks in this limit are expected to exponentiate with the $z$-dependence being subleading \cite{Bonelli:2021uvf,Bonelli_2022}.
They have the asymptotic behavior
\begin{equation}
    \lim_{b\to\infty}\mathfrak{F}(z,\Lambda)\sim e^{b^2\left(F_1(L)+b^{-2}F_2(z,L)+\mathcal{O}(b^{-4})\right)}.
\end{equation} 
The conformal block $F_1(L)$ is related to the irregular conformal block of the 3-point correlator $\langle \epsilon'(0)\epsilon'(1)I_\Lambda(\infty)\rangle$, without the insertion of $\epsilon'$ field at $z$. We denote this conformal block as $\widetilde{\mathfrak{F}}(z,\Lambda)$, which takes the form
\begin{equation}
    \widetilde{\mathfrak{F}}(z,\Lambda)=\Lambda^{\Delta_{01}}e^{b^2\left(F_1(L)+\mathcal{O}(b^{-2})\right)},
\end{equation}
where $\Delta_{01}$ is the conformal weight of the field (the intermediate state $\mathcal{O}_1$ in Figure \ref{fig:4pticb}) from the fusion of $\epsilon'(0)$ and $\epsilon'(1)$ fields.
We may eliminate the divergence and get a finite semiclassical conformal block via normalizing $\mathfrak{F}(z,\Lambda)$ by this block,
\begin{equation}
    \mathcal{F}(z,L)=\lim_{b\to\infty}\frac{\mathfrak{F}(z,\Lambda)}{\widetilde{\mathfrak{F}}(z,\Lambda)}.
\end{equation}
An implication of the subleading behaviour of $z$-dependence of the semiclassical block is that the $\Lambda$-derivative becomes $z$-independent at the leading order, since $\Lambda\partial_\Lambda\mathfrak{F}(z,\Lambda)\sim b^2 \Lambda\partial_\Lambda F_1(\Lambda)\mathfrak{F}(z,\Lambda)$.

Dividing \eqref{eq:irrbpz} by $b^2$, the differential equation for the conformal block $\mathcal{F}(z,L)$ now becomes
\begin{equation}
\begin{aligned}
&\left\{\frac{1}{2}\partial_z^3
    +\frac{2}{z(z-1)}u\partial_z
    +2\left(-\frac{\frac{1}{2}-a_1^2-a_2^2}{z(z-1)}
    +\frac{\frac{1}{4}-a_1^2}{z^2}
    +\frac{\frac{1}{4}-a_2^2}{(z-1)^2}+\frac{mL}{z}-\frac{L^2}{4}\right)\partial_z\right.\\
&\quad \left. +\left( 
         \frac{1-2z}{z^2(z-1)^2}u
         -\frac{1-2z}{z^2(z-1)^2}\left(\frac{1}{2}-a_1^2-a_2^2\right)
         -\frac{\frac{1}{2}-2a_1^2}{z^3}
        -\frac{\frac{1}{2}-2a_2^2}{(z-1)^3}-\frac{mL}{z^2}\right)
\right\}
\mathcal{F}(z,L)=0,
\end{aligned}
\end{equation}
where the parameter $u$ has been introduced as the semiclassical limit of $\Lambda$-derivative:
\begin{equation}
    u :=\lim_{b\to\infty}\frac{1}{b^2}\Lambda\partial_\Lambda \log \widetilde{\mathfrak{F}}(z,\Lambda).
\end{equation}
Thus we have a third-order confluent Heun-like differential equation. Solving explicitly gives us the irregular conformal blocks
\begin{equation}
\begin{aligned}
    \mathcal{F}_1(z,L) 
    &=z^{-\frac{3}{5}} (z-1)^{\frac{13}{5}} e^{L z}  
     \text{HeunC}\left[-\frac{3L}{10} + mL - u, (m+1)L, -\frac{3}{5}, \frac{13}{5}, L, z\right]^2, \\
    \mathcal{F}_2(z,L) &= z (z-1)^{\frac{13}{5}} e^{L z} \text{HeunC}\left[-\frac{3L}{10} + mL - u, (m+1)L, \frac{13}{5}, \frac{13}{5}, L, z\right] \\
    &\quad\times \text{HeunC}\left[-\frac{104}{25} + \frac{13L}{10} + mL - u, (m+\frac{13}{5})L, \frac{12}{5}, \frac{12}{5}, L, z\right], \\
    \mathcal{F}_3(z,L) &=
   z^{\frac{13}{5}} (z-1)^{\frac{13}{5}} e^{L z} \text{HeunC}\left[-\frac{104}{25} + \frac{13L}{10} + mL - u, (m+\frac{13}{5})L, \frac{12}{5}, \frac{12}{5}, L, z\right]^2,
\end{aligned}
\end{equation}
where $\operatorname{HeunC}[q,\alpha,\gamma,\delta,\epsilon,z]$ is the confluent Heun function that satisfies the confluent Heun differential equation 
\begin{equation}
    z(z-1)f''(z)+[\gamma(z-1)+\delta z+ \epsilon z(z-1)]f'(z)+(\alpha z-q)f(z)=0.
\end{equation}
It can be easily verified that for $L=0$, these blocks reduce to those with hypergeometric functions. Similar to the regular block, one can represent pictorially the irregular block as in Figure \ref{fig:4pticb}.

In a similar fashion as the regular blocks, we can expand these blocks around $z=0$, and obtain
\begin{equation}
\begin{aligned}
    \mathcal{F}_1(z,L) \sim z^{-\frac{3}{5}},\quad
    \mathcal{F}_2(z,L) \sim z,\quad
    \mathcal{F}_3(z,L) \sim z^{\frac{12}{5}}.
\end{aligned}
\end{equation}
From these leading exponents, we deduce that the block $\mathcal{F}_1(z,L)$ corresponds to the fusion rule $\epsilon'\times \epsilon'\to \epsilon'$. The other two blocks correspond to fields which lie outside the tricritical Ising model. This is in contrast to the regular case, in which all blocks correspond to fusion rules imposed in the minimal model. We regard this anomaly as the symmetry breaking property of the irregularity in the conformal block. 

\section{Conclusion}
In this paper we have studied the 4-point regular Fibonacci conformal blocks and the corresponding irregular conformal blocks. For the regular case we expressed the blocks as a double integral and obtained their monodromy behavior via the deformation of integration cycles. For irregular blocks we prove that the integral form satisfies a third-order irregular differential equation and obtained the corresponding semi-classical limit. It is promising to obtain the braiding properties of Fibonacci irregular conformal blocks via the same method used on regular blocks.

Apart from applications to quantum computation, there are a number of questions worth further investigation. It was shown that knot and link invariants, namely the Jones polynomial \cite{Witten:1988hf,Gaiotto_2012}, and HOMFLY-PT polynomial as well as Kauffman polynomial \cite{hu2014homfly,Liu_2021}, can be reconstructed from Virasoro conformal blocks. To see whether (and how) the irregular fields modify the above relation between conformal blocks and knot theories is a possible direction to pursue. This is also related to the representations of quantum group and Yang-Baxter equation \cite{Xu:2018ibc,xu2022representations}.

We can also consider more generic conformal blocks, while so far only blocks of degenerate primaries with one rank-1 irregular operator have been studied. The case of higher-rank irregular singularities, double (or more) insertion of irregular operators are yet to be investigated. As we mentioned, Liouville theory has a continuous spectrum of primaries. Conformal blocks of those non-degenerate primaries, though there is no closed form formula, are studied in \cite{Ponsot_2002,ang2023derivation}. To see how the irregular fields fit into this structure is another interesting topic.

\section*{Acknowledgements}
We would like to thank Sergio Cecotti, Sergei Gukov, Yihua Liu, and Nicolai Reshetikhin for valuable discussions. B.H. would also like to thank the Max-Planck Institute for Mathematics in Bonn, where part of this work was completed, for hospitality and financial support. This work was supported by the NSFC grant 12250610187.

\appendix
\section{Irregular BPZ-type equation of level 3}\label{app:irreg}
Consider the irregular correlator $F(z_1,z_2,z_3,z_4,\Lambda)\equiv\langle \mathcal{O}_1(z_1)\mathcal{O}_2(z_2)\mathcal{O}_3(z_3)I_\Lambda(z_4)\rangle$. Instead of acting $L_{-n}$ on the field $I_\Lambda(z_4)$, the adjoint actions $L_n, n>0$ apply accordingly to the properties in \eqref{eq:vgir}. In particular, the action $L_2 I_\Lambda$ will be inserted into the differential equation \eqref{eq:4pde}.
Explicitly, \eqref{eq:4pde} now takes the form:
\begin{equation}
        0=\left[
        \frac{1}{\Delta_1+1}\partial_{z_1}^3
        -2  \sum_{i=2}^3\left(\frac{\Delta_i}{(z_1-z_i)^2}+\frac{1}{z_1-z_i}\partial_{z_i}-\frac{\Lambda^2}{4b^2}\right)\partial_{z_1}
        -\Delta_1\sum_{i=2}^3\left(\frac{2 \Delta_i}{(z_1-z_i)^3}+\frac{1}{(z_1-z_i)^{2}}\partial_{z_i}\right)
        \right]F.
\end{equation}

We eliminate $\partial_{z_2}$ and $\partial_{z_3}$ following the similar procedure as in the regular case. Using the global conformal Ward identities
\begin{equation}
    \begin{aligned}
        &\left(\partial_{z_1}+\partial_{z_2}+\partial_{z_3}-\mu\Lambda\right)F=0,\\
        &\left(z_1\partial_{z_1}+z_2\partial_{z_2}+z_3\partial_{z_3}+\Delta_1+\Delta_2+\Delta_3-\Lambda\partial_\Lambda\right)F=0,\\
    \end{aligned}
\end{equation}
we have
\begin{equation}
\begin{aligned}
    \partial_{z_2}F &=\frac{1}{{z_{23}}}\left[ 
    -(\Delta_1+\Delta_2+\Delta_3)-\mu\Lambda z_3+\Lambda\partial_{\Lambda}-z_{13}\partial_{z_1}\right] F,\\
    \partial_{z_3}F &=-\frac{1}{{z_{23}}}\left[ -(\Delta_1+\Delta_2+\Delta_3)-\mu\Lambda z_2+\Lambda\partial_{\Lambda}-z_{12}\partial_{z_1}\right] F.\\
\end{aligned}
\end{equation}
The parameter $\mu$ here denotes the factor $Q/b=(1+1/b^2)$.

Hence, the differential equation for the irregular correlator reads
\begin{equation}
    \begin{aligned}
        0 &=\left\{\frac{1}{\Delta_1+1}\partial_{z_1}^3
        -2 \left(\frac{\Delta_2}{(z_{12})^2}+\frac{\Delta_3}{(z_{13})^2}-\frac{\Lambda^2}{4b^2}\right)\partial_{z_1}
        -2\Delta_1\left(\frac{\Delta_2}{(z_{12})^3}
        +\frac{\Delta_3}{(z_{13})^3}\right)\right.\\
        &\quad -\frac{2}{z_{12}z_{13}}
        \left[-(\Delta_1+\Delta_2+\Delta_3)+\mu\Lambda(z_1-z_2-z_3)
        +\Lambda\partial_{\Lambda}+(-2z_1+z_2+z_3)\partial_{z_1}\right]\partial_{z_1}\\
        &\quad -\frac{\Delta_1}{(z_{12})^2(z_{13})^2}
        \left[(\Delta_1+\Delta_2+\Delta_3)(-2z_1+z_2+z_3)+\mu\Lambda(z_1^2+z_2^2+z_3^2-2z_1(z_2+z_3)+z_2z_3)\right.\\
        &\qquad \left.\left.-(-2z_1+z_2+z_3)\Lambda\partial_{\Lambda}+(-3z_1^2-z_2^2-z_3^2+3z_1(z_2+z_3)-z_2z_3)\partial_{z_1}\right]\right\}
        F.
    \end{aligned}
\end{equation}
Setting $z_1=z,z_2=0,z_3=1$, we have 
\begin{equation}
    \begin{aligned}
        0&=\left\{\frac{1}{\Delta_1+1}\partial_{z}^3
        -2 \left(\frac{\Delta_2}{z^2}+\frac{\Delta_3}{(z-1)^2}-\frac{\Lambda^2}{4b^2}\right)\partial_{z}
        -2\Delta_1\left(\frac{\Delta_2}{z^3}
        +\frac{\Delta_3}{(z-1)^3}\right)\right.\\
        &\quad -\frac{2}{z(z-1)}
        \left[-(\Delta_1+\Delta_2+\Delta_3)+\mu\Lambda(z-1)^2
        +\Lambda\partial_{\Lambda}+(-2z+1)\partial_{z}\right]\partial_{z}\\
        &\quad \left.-\frac{\Delta_1}{z^2(z-1)^2}
        \left[(\Delta_1+\Delta_2+\Delta_3)(-2z+1)+\mu\Lambda(z-1)
        +(2z-1)\Lambda\partial_{\Lambda}+(-3z^2-1+3z)\partial_{z}\right]\right\}
        F.
    \end{aligned}
\end{equation}
Rearranging the equation gives
\begin{equation}
\begin{aligned}
&\left\{\frac{1}{\Delta_1+1}\partial_z^3
        +2\left( \frac{2z-1}{z(z-1)}\partial_{z}^2
         -\frac{1}{z(z-1)}\Lambda\partial_{\Lambda}\partial_z\right)
         +\Delta_1\frac{1-2z}{z^2(z-1)^2}\Lambda\partial_{\Lambda}\right.\\
&\quad +2\left(\frac{\Delta_1+\Delta_2+\Delta_3}{z(z-1)}
        -\frac{\Delta_2}{z^2}
        -\frac{\Delta_3}{(z-1)^2}
        -\Delta_1 \frac{3z(z-1)+1}{z^2(z-1)^2}
        -\frac{\mu\Lambda}{z}
        +\frac{\Lambda^2}{4b^2}\right)\partial_z\\
&\quad \left.-\Delta_1 \left( 
         \left(\Delta_1+\Delta_2+\Delta_3\right)\frac{1-2z}{z^2(z-1)^2}
        +\frac{2\Delta_2}{z^3}
        +\frac{2\Delta_3}{(z-1)^3}
        +\frac{\mu\Lambda}{z^2}\right)
\right\}\langle \mathcal{O}_1(z_1)\mathcal{O}_2(z_2)\mathcal{O}_3(z_3)I_\Lambda(z_4)\rangle=0.
\end{aligned}
\end{equation}
This is the BPZ-type equation for a 4-point correlator involving a level-3 degenerate primary and an irregular rank 1 vertex operator.

\bibliographystyle{JHEP}     
{\small{\bibliography{main}}}

\end{document}